\documentclass[pdflatex,default]{sn-jnl}

\usepackage{graphicx}%
\usepackage{multirow}%
\usepackage{amsmath,amssymb,amsfonts}%
\usepackage{amsthm}%
\usepackage{mathrsfs}%
\usepackage[title]{appendix}%
\usepackage{xcolor}%
\usepackage{textcomp}%
\usepackage{manyfoot}%
\usepackage{booktabs}%
\usepackage{algorithm}%
\usepackage{algorithmicx}%
\usepackage{algpseudocode}%
\usepackage{listings}%

\theoremstyle{thmstyleone}%
\theoremstyle{thmstyletwo}%

\theoremstyle{thmstylethree}%

\begin{document}

\title[Article Title]{Control of particle transport driven by active noise: strategy of amplification via a periodic potential}


\author[1]{\fnm{Karol} \sur{Bia{\l}as}}

\author[1]{\fnm{Jerzy} \sur{{\L}uczka}}

\author*[1]{\fnm{Jakub} \sur{Spiechowicz}}\email{jakub.spiechowicz@us.edu.pl}


\affil*[1]{\orgdiv{Institute of Physics}, \orgname{University of Silesia}, \linebreak \orgaddress{\street{75 Pu{\l}ku Piechoty 1A}, \city{Chorz{\'o}w}, \postcode{41-500}, \country{Poland}}}




\abstract{We extend our previous studies on a counter-intuitive effect in which a directed transport of a free Brownian particle induced by active fluctuations can be significantly enhanced when the particle is placed in a periodic potential. It is in clear contrast to a common situation when the velocity of the Brownian particle  is notably reduced if the periodic potential is switched on. As a model of active fluctuations we employ white Poissonian shot noise. We reconsider the case of the skew-normal amplitude distribution of shot noise and focus on the impact of statistical characteristics of its amplitude like mean, variance and skewness on the magnitude of free particle transport enhancement. in particular, we detect intriguing oscillations of the rescaled velocity of the particle as a function of the variance. Our findings can be corroborated experimentally in both biological and artificial microscopic systems.}





\maketitle

\section{Introduction}\label{sec1}
An overdamped Brownian particle subjected to a constant force $f$ can be described by the following simple rescaled Langevin equation
\begin{equation}
	\dot{x} = f + \sqrt{2 D_T}\xi(t)
\end{equation}
where $D_T$ corresponds to dimensionless temperature of the system and $\xi(t)$ represents thermal fluctuations modeled by white Gaussian noise of zero mean $\langle \xi(t) \rangle=0$ and correlation function $\langle\xi(t)\xi(s)\rangle=\delta(t-s)$. The mean velocity of the Brownian particle reads
\begin{equation}
	\langle \dot{x}(t) \rangle =  f \equiv v_0.
\end{equation}
where $\langle \cdot \rangle$ stands for the average over noise realizations. Let the system be additionally exposed to a spatially periodic potential $U(x) = U(x+L)$, i.e., 
\begin{equation}
\label{U+f}
	\dot{x} = -U'(x) + f + \sqrt{2 D_T}\xi(t). 
\end{equation}
For a weak constant force, i.e. for $f < \mbox{max}|U'(x)|$, the stationary averaged velocity 
$\langle v \rangle = \lim_{t\to \infty} \langle \dot{x}(t) \rangle$ is notably reduced,  \mbox{$\langle v \rangle \ll v_0$},  due to existence of the potential barriers \cite{risken}. 

Surprisingly, recently \cite{praca_w_PRE,mechanism} it has been reported that when the free particle transport is induced by active nonequilibrium fluctuations $\eta(t)$ with equal statistical bias $\langle \eta(t) \rangle = f$, namely,  
\begin{equation}
	\dot{x} = -U'(x) + \eta(t) + \sqrt{2 D_T}\,\xi(t)
	\label{eq:1}
\end{equation}
transport can be enormously boosted when the particle is additionally placed in a periodic potential, i.e. $\langle v \rangle \gg v_0$. In this work we extend previous study \cite{praca_w_PRE,mechanism} and perform deeper analysis of  the impact of statistical parameters characterizing active nonequlibrium fluctuations $\eta(t)$ on this intriguing effect. In doing so, without loss of generality, we restrict ourselves to the simple spatially symmetric form of the periodic potential
\begin{equation}
    U(x)= -\varepsilon \cos{x}, 
\end{equation}
where $\varepsilon$ represents half of the potential barrier height.

The paper is organized as follows. In the next section we detail on the model of active fluctuations. Then, in Sec. 3, we present a phenomenological derivation of an approximate expression for transport enhancement and focus on the impact of the statistical parameters of fluctuations on the effect of free transport enhancement in the periodic potential. The last section provides brief summary and conclusions.

\section{Model of active fluctuations}\label{sec2}
As a model of active nonequilibrium fluctuations $\eta(t)$ we consider white Poisson shot noise \cite{hanggi1980,spiechowicz2014pre,bialas2020}
\begin{equation}
    \eta(t)=\sum_{i=1}^{n(t)}z_i\delta(t-t_i),
    \label{noise}
\end{equation}
where $t_i$ are arrival times of Poisson point process $n(t)$ \cite{feller1970}, i.e. the probability for occurrence of $k$ impulses in the time interval $[0,t]$ is
\begin{equation}
    Pr\{n(t)=k\}=\frac{(\lambda t)^k}{k!}e^{-\lambda t}.
\end{equation}
The parameter $\lambda$ describes the mean number of impulses per unit time, and consequently its inverse $1/\lambda=\tau_P$ corresponds to the average time between them. The amplitudes $\{z_i\}$ are independent random variables distributed according to  the same probability density function $\rho(z)$. Minimal conditions for  $\rho(z)$ to generate the giant enhancement of the  particle transport have been reported in literature \cite{praca_w_PRE}. If the periodic potential $U(x)$ is symmetric  the probability density $\rho(z)$ has to be asymmetric and possesses variance $\sigma^2=\langle\left( z_i-\zeta\right)^2\rangle$ which is  independent on  the mean $\langle z_i \rangle = \zeta$. Moreover, the support of $\rho(z)$ has to include both positive and negative values. Such a  model can describe both an active particle self-propelling itself inside a passive medium or a passive system immersed in an active bath formed as a suspension of active particles \cite{romanczuk,marchetti,bechinger,dabelow,kanazawa2020,lee2022}. The above constraints are satisfied by e.g.  the skew-normal statistics \cite{azz,rijal2022,bailey2021} defined by the probability density 
\begin{equation}
	\label{rho}
    \rho(z) = \frac{1}{\pi \sqrt{2\pi \omega^2}}e^{-(z-\mu)^2/2\omega^2} \int_{-\infty}^{\alpha[(z-\mu)/\omega]}  \, e^{-s^2/2} \, ds,
\end{equation}
where $\mu$, $\omega$ and $\alpha$ are the location, scale and shape parameters, respectively. These quantities can be redefined in terms of statistical moments of the distribution, i.e. its mean $\zeta$, variance $\sigma^2$ and skewness $\chi=\langle\left(z_i-\zeta\right)^3\rangle/\sigma^3$ \cite{generacja,generacja2}, namely, 
\begin{equation}
\alpha =\frac{\delta}{\sqrt{1-\delta^2}}, \quad
\omega=\sqrt{\frac{\sigma^2}{1- 2\delta^2/\pi}}, \quad
\mu=\zeta-\delta\sqrt{\frac{2\sigma^2}{\pi(1-2\delta^2/\pi)}},
\label{eq_S_def}
\end{equation}
where $\delta$ reads
\begin{equation}
    \delta=\text{sgn}(\chi)\sqrt{\frac{|\chi|^{2/3}}{(2/\pi)\{[(4-\pi)/2]^{2/3}+|\chi|^{2/3}\}}}.
    \label{eq_S_delta}
\end{equation}
For such a choice of parametrization the statistical bias of active fluctuations reads
\begin{equation}
    \langle \eta(t) \rangle =\lambda \langle z_i \rangle =  \lambda \zeta = v_0.
\end{equation}
For comparison,  the deterministic bias is $f=v_0$ as in Eq. (\ref{U+f}). 

\section{Results}\label{sec3}
The integro-differential equation of the Fokker-Planck-Kolmorogov-Feller type corresponding to Eq. (\ref{eq:1}) reads
\begin{equation}
       \frac{\partial P}{\partial t}=\frac{\partial}{\partial x}\left[ U'(x)P(x,t)\right]+D_T\frac{\partial^2}{\partial x^2}P(x,t)+\lambda\int^{\infty}_{-\infty}[P(x-z,t)-P(x,t)]\rho(z) dz.
       \label{eq:fp}
\end{equation}
Unfortunately, generally it cannot be solved analytically \cite{hanggi1980}, in particular for the skew normal amplitude distribution (\ref{rho}). It was achieved only for selected, much simpler special cases \cite{LuczkaBartussek,kanazawa2015,Talbot}. Therefore, we had to  employ precise numerical simulations of the underlying dynamics. In doing so we exploit parallel computing capabilities of graphical processing units that allow  to speed up simulations several orders of magnitude as compared to usual  methods \cite{spiechowicz2015cpc}. The quantity of interest, namely, the rescaled long time velocity $\langle v\rangle/v_0$ of the Brownian particle, is averaged over the ensemble of $2^{16}$  trajectories each starting from various initial conditions for the particle coordinate $x(0)$ distributed uniformly $[0,L]$ over the spatial period $L$ of the potential $U(x)$.
\begin{figure}[t]
    \centering
    \includegraphics[width=0.66\linewidth]{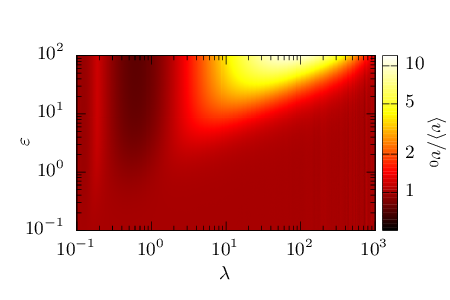}
    \caption{The rescaled velocity $\langle v\rangle/v_0$ (color-coded) of the Brownian particle versus the spiking rate $\lambda$ of active fluctuations and the periodic potential barrier height $\varepsilon$ is depicted for the fixed mean $\langle\eta(t)\rangle = v_0 = \lambda \zeta = 1$ with variance $\sigma^2=3.1$, skewness $\chi=0.99$ and thermal noise intensity $D_T=0.01$.}
    \label{fig:1}
\end{figure}

We start our investigation with the map of rescaled velocity $\langle v \rangle/v_0$ versus the spiking rate $\lambda$ of active fluctuations $\eta(t)$ and the periodic potential barrier height $\varepsilon$,  see Fig. \ref{fig:1}. Please note that along with $\lambda$, the mean amplitude $\zeta$ is varied in order to satisfy the condition $\langle\eta(t)\rangle= v_0 = \lambda\zeta = 1$. The barrier height $\varepsilon$ is related to the mean relaxation time $\tau_R \propto 1/\varepsilon$ of the particle towards its minimum \cite{praca_w_PRE,mechanism}. On the other hand, the spiking rate $\lambda$ determines the average time $\tau_P = 1/\lambda$ between two successive $\delta$-impulses of active fluctuations. A careful inspection of Fig. \ref{fig:1} allows us to observe that the relation between these two characteristic time scales $\tau_R$ and $\tau_P$ is a decisive factor for  magnitude of the rescaled velocity $\langle v \rangle/v_0$. It turns out that for the fixed statistical bias $\langle \eta(t) \rangle = v_0 = const$ the optimal amplification $\langle v \rangle/v_0 > 1$ of the  particle transport occurs in the resonance regime in which $\tau_R \approx \tau_P$ \cite{praca_w_PRE,mechanism}. On average, after each $\delta$-impulse  the particle relaxes by the potential gradient to the vicinity of its minimum and next due to optimized Poisson jumps it overcomes the potential barrier relaxing to the nearest minimum.  Therefore the particle is able to fully exploit the potential to enhance its velocity. As we can see the larger the barrier height $\varepsilon$, the greater the magnitude of transport enhancement as well as the interval of $\lambda$ in which this effect emerges. On the other hand, if the spiking rate $\lambda$ is increased the minimal barrier height $\varepsilon$ needed for the free particle transport amplification grows as well. This finding follows from the resonance condition $\tau_R \approx \tau_P$ for which the transport boost is optimal. If the barrier height $\varepsilon$ is too small this effect does not survive since then the statistical bias $\langle \eta(t) \rangle > \varepsilon$ and impact of the periodic potential becomes negligible.
\begin{figure}[t]
    \centering
    \includegraphics[width=0.66\linewidth]{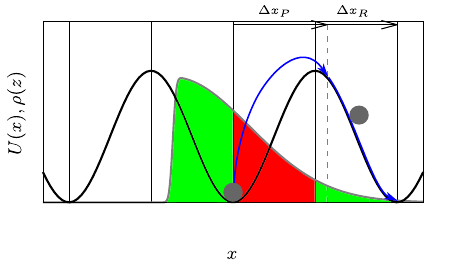}
    \caption{Schematic representation of the jump-relaxation process as a phenomenological description of the system dynamics in the absence of thermal fluctuations and with the exemplary amplitude distribution $\rho(z)$ of  mean amplitude $\zeta=1/30$, variance $\sigma^2=3.1$ and skewness $\chi=0.99$. Green areas under the amplitude distribution correspond to the relaxation in the right direction  $\Delta x_R(\tau,\Delta x_P)>0$ and the red one to the left direction $\Delta x_R(\tau,\Delta x_P)<0$.}
    \label{fig:2}
\end{figure}

Although the Fokker-Planck Eq. (\ref{eq:fp}) cannot be solved in an analytic way, a phenomenological expression can be inferred to describe the spatial coupling between nonequilibrium noise amplitude distribution $\rho(z)$ and the periodic potential $U(x)$. Let us consider a process presented in \mbox{Fig. \ref{fig:2}}: a particle at the initial position $x_0$ in the potential minimum in the absence of thermal fluctuations. To simplify the forthcoming formulas, we can safely assume that $x_0=0$. 
At the minimum the particle experiences a kick induced by the Poisson process $\eta(t)$ and moves in the right direction over the distance $\Delta x_P$. Next it relaxes towards the potential minimum and travels the distance $\Delta x_R = \Delta x_R(\tau,\Delta x_P)$. Here, $\tau$ is a random time interval between two consecutive $\delta$-impulses whose probability density function $\varphi(\tau)$ reads
\begin{equation}
    \varphi(\tau)=\theta(\tau) \lambda e^{-\lambda \tau}.
\end{equation}
where $\theta(\tau)$ is the Heaviside step function.  

The expression for $\Delta x_R(\tau,\Delta x_P)$ can be obtained by solving the noiseless, deterministic differential equation $\dot{x}=-U'(x)$. It reads
\begin{equation}
    \Delta x_R(\tau,\Delta x_P)=2\arctan \left[\tan\left(\frac{y(\Delta x_P)}{2}\right)e^{-\varepsilon\tau}\right]-y(\Delta x_P). 
\end{equation}
In order to invert the function tan(x), we have repositioned $\Delta x_P$ to the interval $(-\pi,\pi)$ by making use of the transformation $y(\Delta x_P)=[ (\pi+\Delta x_P)\mod{2\pi} ] -\pi$. 
The average velocity of the particle is
\begin{equation}
    \langle v\rangle=\frac{\langle \Delta x\rangle}{\tau_P}=\frac{\langle \Delta x_P+\Delta x_R \rangle}{\tau_P}=\lambda (\langle \Delta x_P\rangle+\langle\Delta x_R\rangle),
    \label{eq:v_r}
\end{equation}
Consequently, the rescaled version reads
\begin{equation}
    \frac{\langle v\rangle}{v_0}=\frac{\lambda (\zeta+\langle\Delta x_R\rangle)}{\lambda \zeta}=1+\frac{\langle \Delta x_R\rangle}{\zeta}.
    \label{eq:v_en_r}
\end{equation}
\begin{figure}[t]
    \centering
    \includegraphics[width=0.66\linewidth]{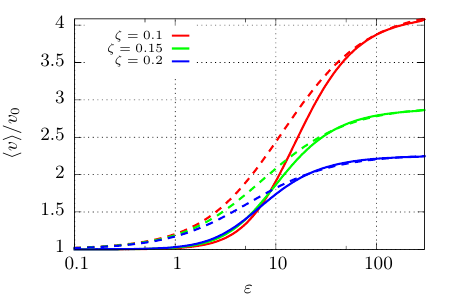}
    \caption{The rescaled velocity $\langle v\rangle/v_0$ of the Brownian particle versus the periodic potential barrier height $\varepsilon$ is depicted for different mean amplitudes $\zeta$ with fixed mean bias $\langle\eta(t)\rangle = v_0 = \lambda \zeta = 1$, variance $\sigma^2=3.1$, skewness $\chi=0.99$ and thermal noise intensity $D_T=0.01$. Solid lines represent values obtained from precise numerical simulations of Eq. (\ref{eq:1}) whereas the dashed ones are calculated using the phenomenological expression in Eq. (\ref{eq:v_en_r}).}
    \label{fig:3}
\end{figure}
Here  $\langle \Delta x_P\rangle$ is just the  mean amplitude $\langle z_i \rangle = \zeta$ distributed according to the probability  density  $\rho(z)$. On the other hand, $\langle\Delta x_R\rangle$ can be represented as the  particle displacement $\Delta x_R(\tau,\Delta x_P)$ due to the relaxation towards the potential minimum averaged over the corresponding distributions for the random time $\tau$ between two consecutive $\delta$-spikes of Poisson noise and their amplitudes, namely, 
\begin{equation}
\label{final}
    \langle\Delta x_R\rangle=\int_{-\infty}^{\infty}\int_0^{\infty}\varphi(\tau) \rho(z)\Delta x_R(\tau,z)d\tau dz.
\end{equation}
From Eq. (\ref{eq:v_en_r}) we can infer that when $\langle\Delta x_R\rangle\to 0$, e.g. when spiking rate is very high, the rescaled velocity $\langle v\rangle/v_0$ tends to unity. It is because the particle does not have sufficient time to exploit the relaxation process before another $\delta$-spike strikes it. This limit is also satisfied when the potential barrier height is very small.
In Fig. \ref{fig:3} we compare the results obtained from precise numerical simulations of Eq. (\ref{eq:1}) and  those derived from Eqs. (\ref{eq:v_en_r})-(\ref{final}). The latter is correct in two extreme regimes of small and large potential barrier $\varepsilon$. In the first case 
$\tau_P\ll\tau_R$ and enhancement does not emerge due to the fact that $\langle\Delta x_R\rangle\to 0$ whereas in the second situation $\tau_P\gg\tau_R$ and the particle always relaxes to the vicinity of the potential minimum. The proposed expression is not accurate for the regime of the moderate potential barrier corresponding to the resonance condition $\tau_P\approx\tau_R$. It is a consequence of the assumption that initially the particle resides at the potential minimum, see Fig. \ref{fig:2}. This condition in such a regime is only satisfied \emph{on average} while full dynamics of the system is still random. In particular, nonequilibrium noise impulses arriving far from the potential minimum disrupt our phenomenological description and the average velocity calculated from the precise numerical simulations of the full dynamics is lower than the one from the jump-relaxation process. Since a detailed discussion on the mechanism of the studied effect is presented elsewhere \cite{mechanism} we now turn to the main goal of our work which is a complementary analysis of the influence of parameters characterizing the skew-normal statistics $\rho(z)$ of active fluctuation amplitudes on the directed transport.
\begin{figure}[t]
    \centering
    \includegraphics[width=0.66\textwidth]{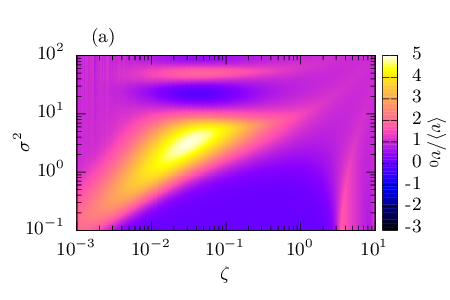} \\
    \includegraphics[width=0.66\textwidth]{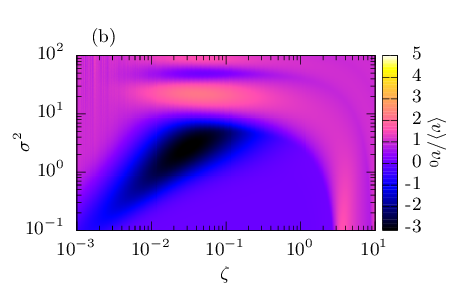}
    \caption{The rescaled velocity $\langle v\rangle/v_0$ (color-coded) as a function of mean amplitude $\zeta$ and variance $\sigma^2$ of active fluctuations $\eta(t)$ with fixed bias $\langle \eta(t) \rangle = v_0 = 1$. Other parameters are: skewness $\chi=0.99$ (panel (a)), $\chi=-0.99$ ( panel (b)), the barrier height $\varepsilon=40$ and thermal noise intensity $D_T=0.01$.}
    \label{fig:4}
\end{figure}
In Fig. \ref{fig:4} we show the rescaled velocity $\langle v \rangle/v_0$ as a function of mean amplitude $\zeta$ and variance $\sigma^2$ of the distribution $\rho(z)$ for the positive $\chi = 0.99$ and negative $\chi = -0.99$ skewness. The reader can immediately notice that the most radical change in the particle velocity occurs for small mean $\zeta \ll 1$ and for variance $\sigma^2$ of the order of the potential period $\sigma^2 \approx L$. There are two reasons for this behavior. The first is the relation  $\tau_R \approx \tau_P$ between the characteristic time scales of the particle relaxation $\tau_R$ and the average interval $\tau_P$ separating two successive impulses of active fluctuations. The periodic potential can significantly boost the force-free particle transport when its impact is meaningful, i.e. magnitude of the barrier $\varepsilon$ is considerable. It implies that in order to maintain the resonance regime $\tau_R \approx \tau_P$ the spiking rate $\lambda$ must increase as well. This in turn means that the mean amplitude $\zeta$ has to  decrease in order to satisfy the condition $\langle \eta(t) \rangle = v_0 = \lambda \zeta = const.$. The second reason is the spatial coupling between the amplitude statistics $\rho(z)$ and the periodic potential $U(x)$. The former cannot be too extended and when the particle is kicked by the $\delta$-spike it should promote the barrier crossing events in the direction indicated by the statistical bias $\langle \eta(t) \rangle = v_0$ rather than in the opposite one. If the first case takes place the free particle transport velocity is boosted $\langle v \rangle > v_0$ whereas the second scenario leads to reversal of the current $\langle v \rangle < 0$.
\begin{figure*}[t]
    \centering
    \includegraphics[width=0.49\textwidth]{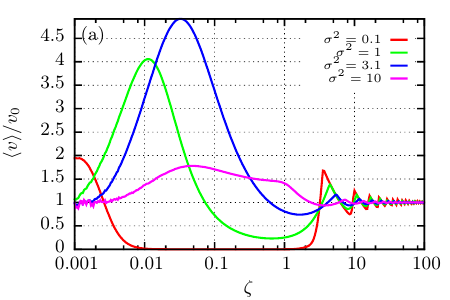}
    \includegraphics[width=0.49\textwidth]{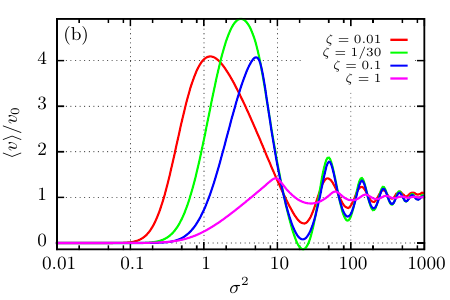}
    \includegraphics[width=0.49\textwidth]{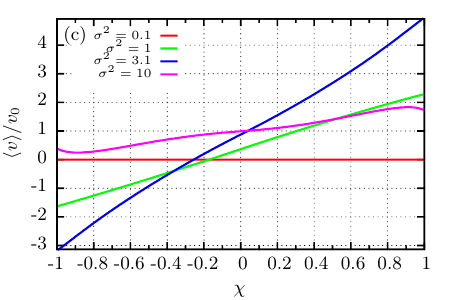}
    \caption{The rescaled velocity $\langle v\rangle/v_0$ as a function of different parameters characterizing active fluctuations amplitude statistics $\rho(z)$: panel (a) mean amplitude $\zeta$, (b) variance $\sigma^2$ and (c) skewness $\chi$. The statistical bias is fixed to $\langle \eta(t) \rangle = v_0 = 1$. Other parameters are the barrier height $\varepsilon = 40$, the spiking frequency $\lambda = 30$ and thermal noise intensity $D_T = 0.01$. In panel (a) and (b) the skewness reads $\chi = 0.99$.}
    \label{fig:5}
\end{figure*}
The above mentioned spatial coupling between the amplitude statistics $\rho(z)$ and the periodic structure $U(x)$ is also visible in two oscillatory regimes where the average velocity $\langle v \rangle$ oscillates around the free particle transport $v_0$ which we present in \mbox{Fig. \ref{fig:5}}. The first one is shown in panel (a) where we depict the rescaled velocity $\langle v \rangle/v_0$ as a function of the mean amplitude $\zeta$ for different values of variance $\sigma^2$ and skewness $\chi = 0.99$. It is the most pronounced for small variance, see the case $\sigma^2 = 0.1$, when the amplitude distribution is very compact. As the mean amplitude $\zeta$ is increased and crosses multiples of the distance $L/2$ between the potential minimum and maximum, transport is alternately greater and lesser than the free particle velocity. When it is boosted $\delta$-spikes move the particle over the potential barrier on average and it relaxes forward towards the next minimum. If it is hampered they are not able to achieve this goal statistically and the particle relaxes backward towards the minimum where it waits for the arrival of $\delta$-spike. Nevertheless, this regime is not very interesting since impulses with small variance, i.e. almost constant amplitude, are not a very good model for active fluctuations. The second type of oscillations occurs when $\zeta$ is small and variance $\sigma^2$ is increased, see panel (b) of the same figure. The mechanism of oscillations is similar since the only difference is that in such a case the amplitude distribution $\rho(z)$ covers more than one spatial period of the potential. If the variance $\sigma^2$ is sufficiently large the impact of the periodic substrate becomes negligible and transport tends to the free particle velocity.

In order to get the full picture of how amplitude statistics parameters influence this effect,  in panel (c) we present the rescaled velocity $\langle v \rangle/v_0$ as a function of the skewness $\chi$ for selected values of variance $\sigma^2$. The reader can observe that change in the distribution asymmetry $\chi$ can reverse the direction of particle current and the latter is still boosted in comparison to the free particle transport. Moreover, the relation between the rescaled velocity $\langle v \rangle/v_0$ and skewness $\chi$ is almost linear for moderate variance $\sigma^2$.

\section{Conclusions}\label{sec4}
In summary, in this work we introduced the phenomenological approximate expression for velocity and investigated the impact of parameters characterizing statistical properties of the amplitude distribution of active fluctuations on transport of an overdamped Brownian particle in the periodic potential. We have demonstrated that within tailored parameter regimes of the mean amplitude $\zeta$, variance $\sigma^2$ and skewness $\chi$, the force-free directed transport can be significantly boosted when the particle is subjected to the periodic  potential. In particular, it turns out that this effect is most pronounced for small $\zeta$, moderate $\sigma^2$ and large $\chi$. Moreover, we have detected intriguing oscillations of the particle rescaled velocity as a function of the variance of active fluctuations amplitude  which is related to forward and backward relaxation in the periodic potential. Our findings can be corroborated experimentally in both biological systems \cite{ezber,ariga} immersed \textit{in situ} in sea of thermal and active fluctuations or in artificial microscopic setups, e.g. Josephson junctions \cite{spiechowicz2015chaos} or colloidal particles in optical potentials \cite{park,paneru}.

\section*{Acknowledgments}
This work has been supported by the Grant NCN No. 2022/45/B/ST3/02619 (J.S.) and by the funds granted under the Research Excellence Initiative of the University of Silesia in Katowice.

\section*{Data availability statement}
The data that support the findings of this study are available from the corresponding author upon request.


\end{document}